# Science Advances

**Manuscript Template**

## FRONT MATTER

### Title

- Ultra-high Purcell factor using long-range modes in asymmetric plasmonic waveguides

### Authors

Y. Su,[1] P. Chang,[1] C. Lin,[1] A. S. Helmy[1]*

### Affiliations

[1]Electrical and Computer Engineering, University of Toronto, Toronto, Canada

*a.helmy@utoronto.ca

### Abstract

For integrated optical devices, realistic utilization of the superior wave-matter interaction offered by plasmonics is typically impeded by optical losses, which increase rapidly with mode volume reduction. Although coupled-mode plasmonic structures has demonstrated effective alleviation of the loss-confinement trade-off, stringent symmetry requirements must be enforced for such reduction to prevail. In this work, we report an asymmetric plasmonic waveguide that is not only capable of guiding subwavelength optical mode with long-range propagation, but is also unrestricted by structural, material, or modal symmetry. In these composite hybrid plasmonic waveguides (CHPWs), the versatility afforded by coupling dissimilar plasmonic structures, within the same waveguide, allow better fabrication tolerance and provide more degrees of design freedom to simultaneously optimize various device attributes. The CHPWs used to demonstrated the concept in this work, exhibit propagation loss and mode area of only 0.03 dB/μm and 0.002 μm$^2$ respectively, corresponding to the smallest combination amongst experimentally demonstrated long-range plasmonic structures to-date. Moreover, CHPW micro-rings were realized with record in/out coupling excitation efficiency (71%), extinction ratio (29 dB), as well as Purcell factor ($1.5 \times 10^4$).

## MAIN TEXT

### Introduction

In plasmonic structures, optical modes experience enhanced field confinement and density of states unattainable within diffraction-limited dielectric counterparts (1 – 3). These attributes allow unprecedented integration density as well as enhancement of linear and nonlinear optical processes (4), which inspired the development of plasmon-based devices for integrated nano-optics, signal processing, spectroscopy, and sensing (5 – 7). Nonetheless, the wave-matter interaction enabled by plasmonics deteriorates with optical losses, which increase rapidly with field localization (8, 9). As such, effective performance improvement is only observed for inherently weak physical processes such as fluorescence and Raman scattering (10), which typically take place over extended surfaces. For subwavelength, guided-wave plasmonic devices, particularly ones that require strong feedback mechanism and Purcell factor, performance attributes such as efficiency and sensitivity are still sub-optimal due to significant insertion loss (11 – 14).

To-date, numerous strategies have been proposed to alleviate the loss-confinement trade-off for guided-wave plasmonic structures. One approach utilizes gain media to compensate for modal losses, but has seen limited success due to the high current densities required (15). The second effort focuses on reducing the field penetration into the metal region via multi-dielectric hybrid plasmonic structures (16, 17). Recently, a coupled-mode plasmonic waveguide was proposed, which utilizes the destructive interference between two hybrid plasmonic modes to achieve subwavelength supermode with propagation loss akin to long range plasmonic structures (18).

In this work, we report a new class of plasmonic structures named composite hybrid plasmonic waveguides (CPHW). It is demonstrated, for the first time, that the destructive interference between dissimilar plasmonic modes can also minimize loss without sacrificing modal area. Unlike previous long-range plasmonic structures that are constrained by structural, material, or modal symmetry, the asymmetrical CHPW structure provides additional design parameters to tailor the electromagnetic field distribution for device performance enhancement. Specifically, ultra-compact plasmonic rings were experimental demonstrated with record propagation loss, excitation efficiency, extinction ratio, as well as Purcell factor.

**Results**

The CHPW structure is designed to be asymmetric with respect to the metal layer as schematically shown in Fig. 1(a). The bottom metal-insulator-semiconductor stack supports a hybrid plasmonic mode (HPW) while the top metal-semiconductor interface supports a surface plasmon polariton mode (SPP). For metal thickness less than the skin depth, the two dissimilar plasmonic modes couples together to form transverse-magnetic (TM) plasmonic supermodes. Moreover, for sufficiently narrow waveguide width, any transverse-electric modes become cut-off. Detailed analysis of the guided modes can be found in the supplementary material.

The propagation loss of the TM supermodes are primarily determined by the field overlap between the longitudinal field component ($E_z$) and the metal layer (19). The long-range (LR) and short-range (SR) supermodes correspond to destructive and constructive interference of $E_z$ respectively. As such, the LR supermode can have lower modal loss due to reduced field overlap. Specifically, loss can be reduced by minimizing the longitudinal electric flux inside the metal layer, as described by $\Phi_{mz} = \int_{A_m} \text{Im}\{E_z\} / P_z \, A$, where $P_z$ is

Poynting vector and the integration is performed over the metal layer of cross-section area $A_m$. For 1D plasmonic stack, $\Phi_{mz}$ can be reduced to zero if $E_z = 0$ at the center of the metal layer, corresponding to complete anti-symmetric $E_z$ distribution. For 2D waveguides with finite width, a fully anti-symmetric field cannot exist due to fringe effects (Suppl. Mat. Fig. S4). Hence, $E_z$ cannot be zero at the center of the metal over the entire waveguide width. Nonetheless, reducing $\Phi_{mz}$ within the entire metal cross-section remains an effective strategy towards loss reduction. As an example, $\Phi_{mz}$ of the LR supermode is tuned by varying the thickness of the top high-index dielectric layer ($t_{\varepsilon 4}$), which is chosen as it can be well-controlled in fabrication (Fig. 1(c)). Despite the different guiding nature and hence field distribution of the HPW and SPP modes, $\Phi_{mz}$ can be drastically tuned and in this example is reduced from 1.03 to 0.003 V·m/W as $t_{\varepsilon 4}$ is changed from 170 to 185nm (Fig. 1(b)). Such a substantial reduction in turn leads to a minimal CHPW loss of only 0.02 dB/μm at $t_{\varepsilon 4} = 185$nm, a record for plasmonic modes with subwavelength modal confinement. Moreover, this corresponds to 20 and 750 times reduction compared to the losses of a standalone HPW and SPP modes, which are the constituents of this CHPW, respectively.

The effective mode area of the LR supermode is shown in Fig. 1(c). It is observed that the mode area increases rapidly as thickness is decreased due to field leaking into the air cladding while it remains relatively constant at larger thicknesses as most of the field is localized in the low-index layer. The effective mode area is 0.002 μm$^2$ at $t_{\varepsilon4}$ = 185nm and changes by only 5% if the optimal $t_{\varepsilon4}$ varies by 10%.

Independent of the modal asymmetry, flux minimization is a robust and novel strategy that lowers the propagation loss of the CHPW while maintaining its strong confinement. The ability to combine different plasmonic modes at a single metal interface allows arbitrary structures and material to be utilized, which in turn provides more degrees of freedom to not only better balance the loss-confinement trade-off compared to previous long-range plasmonic waveguide designs, but also better tailor the field distribution to improve specific device performance attributes (18, 19). Note that the TE cut-off condition is only weakly influenced by $t_{\varepsilon4}$ and the low-loss condition also does not depend strongly on wavelength (Suppl. Mat. Fig. S5) further enhancing the fabrication tolerance of this technique. The strategy of reducing the flux within the metal layer, $t_{\varepsilon4}$ can be achieved using any structural parameter in the waveguide, as well as bias-induced changes to the optical properties, which suggest that the minimum loss condition can be achieved post-growth in a dynamic fashion using bias or current. A comparison of the CHPW's figure-of-merit against other plasmonic waveguides can be found in Suppl. Mat. Table SI. To demonstrate the capability of the proposed CHPW structure, CHPW waveguides and rings were fabricated on standard 220nm-thick SOI wafers (Fig. 2(a)). First, the 20nm SiO$_2$ and 10nm Al layers are deposited via plasma enhanced chemical-vapor deposition and sputtering respectively. Next, α-Si is sputtered and partially etched down to a thickness ranging from 155 to 250nm. Finally, the waveguides are patterned with electron beam lithography and the dielectric and Al layers are etched via deep reactive ion etching and wet processing respectively. CHPWs with lengths ranging from 10 to 400μm were fabricated for cut-back measurements.

The CHPW propagation loss is measured at λ = 1550nm and plotted as function of $t_{\varepsilon4}$ in Fig. 2(c). The measured loss values match well with theoretical predictions, with average loss of 0.03 dB/μm at $t_{\varepsilon4}$ = 185nm. If $t_{\varepsilon4}$ deviates from the minimum flux condition, loss increases to 0.07 and 0.09 dB/μm for $t_{\varepsilon4}$ = 155 and 250nm respectively. The two orders-of-magnitude loss reduction compared to other experimental plasmonic waveguide demonstration is shown in Fig. 2(c) (20).

The utilization of an HPW as part of the coupled-mode structure enables highly efficient, instantaneous CHPW mode excitation without the need for taper structures. Specifically, non-resonant, end-butt excitation using silicon nanowires can be employed as the inclusion of a low-index $\varepsilon_2$ reduces the field distribution and propagation constant mismatch between the two type of waveguides (see Sup Mat Fig. S6). As a result, record excitation efficiency of 56% to 71% is measured at junctions between 200nm-wide CHPW and 800nm-wide silicon nanowire, which compares well when compared to the theoretically predicted range of 65 – 80%.

With minimal modal loss, the strong mode confinement can now facilitate the enhancement of Purcell factor. For proof-of-concept, CHPW ring resonators with various radii and bus-to-ring coupling gap spacing were fabricated and their measured spectra are shown in Fig. 3. Critical coupling is observed at a gap spacing of 270nm, with a record extinction ratio up to 29 dB. The compact radii of ~2.5µm allows for large free spectral range of 44.8nm and average redshift of 48 pm/°C between $30-75\,°C$ was observed (Suppl. Mat. Fig. S12). Given the calculated 0.002 µm$^2$ cross-sectional mode area, a 2.5 µm-radius ring resonator only has an effective mode volume $V_{eff}$ of 0.032 µm$^3$. Therefore, our experimental device can achieve a normalized Purcell factor ($Q/V_{eff}$) as high as $1.5 \times 10^4$, which is three times higher than the state-of-the-art plasmonic-based as well as all-silicon-based micro-ring resonators (Fig. 4) (21 – 24).

**Discussion**

In conclusion, we reported a CHPW architecture where dissimilar plasmonic modes are coupled together at a single metal interface to simultaneously reduce field overlap with lossy metal and confine power within subwavelength areas. Such platform does not require structural or even modal symmetry, therefore allow better fabrication tolerance, liberate long-range plasmonic structures from materials restrictions, and provide more degrees of design freedom to simultaneously optimize various device attributes. The flexibility of the CHPW not only leads to the first experimental demonstration of long-range hybrid plasmonic waveguides, but also plasmonic micro-ring resonators with record loss value and Purcell factor. These plasmonic waveguides will no doubt have impact that extends from ultra-compact interconnects, non-linear sensing, to integrated circuitry that is programmable and reconfigurable (13, 25 – 29). On the other hand, the plasmonic resonators are highly attractive for lasing using a plasmonics platform, where undesired loss due to imperfect reflection at end-facets of Fabry-Perot-like cavities can be avoided (13, 30), as well as plasmon cavity quantum electrodynamics, particularly in the strong coupling regime where Purcell factor is still a limiting factor (31).

**Materials and Methods**

**CHPW fabrication**

We start from a bare standard 220nm SOITEC wafer. Deposition of 20nm-thick SiO$_2$ is carried out using Oxford Instruments PlasmaLab System100 PECVD at a plasma temperature of 400°C using SiH$_4$ and N$_2$O. Deposition of 10nm-thick Al is done using AJA International ATC Orion 8 Sputter Deposition System at room temperature in Ar plasma. Finally, deposition of α-Si:H is performed using MVSystems Multi-Chamber (Cluster Tool) PECVD at a substrate temperature of 180°C using SiH$_4$ plasma. Lithography to define etching patterns is carried out in Vistec EBPG 5000+ Electron Beam Lithography System. As silicon nanowires and CHPW are built on the same substrate, multiple alignment steps are required. Waveguide patterning is realized using Oxford Instruments PlasmaPro Estrelas 100 deep reactive ion etching for silicon and oxide layers, and aluminum etchant type A for the metal layer.

**H2: Supplementary Materials**

Optical waveguides based on surface plasmon polaritons (SPP) offer the capability to propagate and confine light within structural dimensions much smaller than what the diffraction limit allows in conventional methods such as total internal reflection (TIR) (9). SPP are electromagnetic waves that can propagate along a metal-dielectric interface, via coupling of the electromagnetic fields to oscillations of the electron plasma near the surface of a conductor (6). In contrast to TIR, SPP are surface modes with evanescent fields perpendicular to the interface and decay exponentially into the metal and dielectric, with the highest energy localization close to the interface. Plasmonic waveguides can confine light on subwavelength scales and allows designing devices with extremely high local light-matter interaction, reducing device footprint and interaction length. However, the propagation losses incurred during the energy exchange between photons and surface plasmons in metals is often a barrier for adoption of plasmonic devices as the propagation length is limited.

To alleviate this limitation, plasmonic waveguides are typically designed in coupled modes to obtain favorable loss-confinement trade-off. An example is metal-insulator-metal (MIM) waveguide which consists of a nanoscale dielectric slot between two metal interfaces and supports the highest mode localization, however at the cost of extremely short propagation length due to absorption losses in the metal (32). On the other hand, insulator-metal-insulator (IMI) waveguides provide longer propagation lengths at the expense of loose confinement (33). To curtail these shortcomings and achieve a more optimized trade-off, the most prominent approach is to sandwich a low-index dielectric between metal and high-index dielectric to form a hybrid plasmonic waveguide (HPW), thereby delocalizing a portion of the mode away from the metal layer to reduce absorption losses (16, 17). Furthermore in Wen et al., it was demonstrated that propagation losses can be further optimized without sacrificing mode confinement in asymmetric hybrid plasmonic waveguides (AHPW) via field symmetry engineering inside the metal layer (18). Based on the methodology applied to design the AHPW for long-range propagation, we designed a novel composite hybrid plasmonic waveguide (CHPW) to achieve record low propagation losses in plasmonic structures while still maintaining mode areas comparable to MIM slot waveguides.

**Composite Hybrid Plasmonic Waveguide - Design & Analysis**

The CHPW shown in Fig. S1 consists of the superposition of a standard HPW mode with an SPP mode coupled through a thin metal film. In this structure, the asymmetry does not only come from permittivity differences of the materials (structural asymmetry), but rather the coupled modes are formed from modes that operate differently at the fundamental level. The combination of HPW and SPP into a single structure suggests light-matter interaction on two types of plasmonic interfaces and allows one waveguide to support multiple applications, more degrees of freedom in engineering the field overlap to enhance the loss-confinement trade-off and flexible design of active devices.

From coupled mode theory, CHPW supermodes are formed from coupling two individual waveguides through a common metal film. When the metal thickness is reduced below its skin depth, coupled supermodes between HPW and SPP arise as the evanescent fields extend to the opposite side and perturb each other, leading to the formation of a plasmonic long-range (LR) TM mode that is symmetric in the $y$-direction and anti-symmetric in the $z$-direction, and short-range (SR) that is anti-symmetric in $y$ and symmetric in $z$. The field profiles of the plasmonic modes are plotted in Fig. S2. The configuration of the CHPW is the same as the one proposed in the main text which consists of $SiO_2$ substrate ($\varepsilon_{sub}$), 220nm Si ($\varepsilon_1$), 20nm $SiO_2$ ($\varepsilon_2$), 10nm Al ($\varepsilon_3$), 147nm Si ($\varepsilon_4$) and air cladding ($\varepsilon_{air}$). It is seen that for the LR supermode, the field profile of $E_y$ is symmetrical-like while $E_z$ is anti-symmetrical-like across the metal film. This is due to the relative phase shift between $E_y$ and $E_z$ in a typical TM waveguide, as the anti-symmetric supermodes exhibits the modal profiles that are complimentary to its symmetric counterpart.

To briefly examine the dispersion properties, the HPW side is kept constant as we change the SPP side layer thickness ($t_{\varepsilon4}$) is varied in Fig. S2. The SR supermode corresponds to in-phase coupling which results in strong field interaction across the common metal layer, leading to significant absorption losses, while the LR supermode is the opposite and minimizes mode overlap with the metal as a result of destructive interference. It should be noted that the anti-crossing behavior commonly observed in coupled systems is not present in Fig. S2(f). This is evidence that the coupled structured has an extremely high degree of asymmetry as the original HPW and SPP modes have fundamentally different dispersion properties.

**CHPW Design Using Flux Engineering**

To obtain the condition for minimization of CHPW propagation loss, we begin by considering a 1D CHPW as schematically shown in Fig. S3. It is seen that by adjusting the thickness of the top silicon layer, the CHPW can reach a minimum point in the propagation loss, which suggests the amount of modal field overlapping with the metal film is minimized. In the present architecture, the optimized propagation loss is achieved when the top thickness is 147nm. As the electric field of plasmonic waveguide within the metal film is predominantly in the propagation direction $z$, it is hence instructive to study the corresponding field distribution ($E_z$) of CHPW inside the metal, depicted in Fig. S3(a).

Notably, when the propagation loss is minimized, the electric field is rendered anti-symmetrically distributed across the metal as the field becomes zero at the center, while the zero-crossing of $E_z$ will shift away from the center when the thickness deviates from 147nm. The position of the zero-crossing of the electric field, resulting from the field cancellation of two dissimilar top and bottom plasmonic waveguides, hence implies the degree of anti-symmetry as well as the amount of modal field that overlaps with the metal layer. In 1D configuration, the anti-symmetric field profile with the zero-crossing being positioned at the center of the metal thus can serve as the design paradigm for optimizing the propagation loss of 1D CHPW. It should be noted that although the field cancellation (zero-crossing) also occurs for other $\varepsilon_4$ thicknesses, it is the location of the zero crossing that dictates the propagation loss of the plasmonic mode. As shown in Fig. S3(b), although the electric field can be rendered very close to zero in the vicinity of the metal-dielectric interface for $t_{\varepsilon4} = 135$ or 165nm, the propagation loss is substantially higher than that of 147nm.

Since the propagation loss is due to the non-zero field that overlaps with the lossy metal film, it can be deduced that the amount of modal field contained in the metal can be minimized when the field profile is anti-symmetric with respect to the center of the metal layer. Specifically, the field anti-symmetry across the metal layer, or the centered zero-crossing, suggests that there is equal amount of positive and negative modal fields within the metal. In 1D configuration, the exact balance of these opposite sign fields in turn give rise to zero total $E_z$ field in the metal layer, given by:

$$f_{mz} = \int_{t_m} \frac{\mathrm{Im}\{E_z\}}{P_z} \, l \quad \text{(S1)}$$

where $P_z$ is the Poynting vector in the propagation direction and serves as a normalized factor in the calculation. The integration is performed through a line integral of the imaginary part of $E_z$, which is the predominant component for $E_z$ within the metal layer of thickness $t_m$.

Instructively, one thus can use Eq. S1, which is the integration of the electric field over the whole metal layer, to quantify the amount of modal field overlaps with the metal. In other words, Eq. S1 offers a design metric that can characterize the degree of anti-symmetry or the position of zero crossing of $E_z$ for 1D CHPW. For a given metal thickness, the propagation loss thus can be optimized by suitably engineering the modal $E_z$ field resident in the metal layer. As depicted in Fig. S3(b), $f_{mz}$ varies sensitively with the thickness of the top silicon layer, and is rendered zero when the loss of CHPW becomes minimal.

## 2D CHPW

While analysis of the mode confinement in 2D is necessary for design of devices and circuits, the mode profiles and field symmetry show a similar pattern to the 1D case. To completely cut-off the unwanted TIR TE supermodes, the CHPW width is narrowed down to 200nm as seen in Fig. S5(a). However, due to fringe effects from finite width and leakage into outer layers due to diffraction, the $\varepsilon_4$ thickness needed to achieve optimized field symmetry varies from the 1D case.

By extension from the 1D design, the minimization of the total $E_z$ field can also become a design paradigm for optimizing the loss of 2D CHPW, even though the modal field is not invariant in the lateral direction. The integration in Eq. S1 can thus be extended to 2D structure:

$$\Phi_{mz} = \int_{A_m} \frac{\mathrm{Im}\{E_z\}}{P_z} \, A \quad (\mathrm{S2})$$

where $P_z$ is the Poynting vector of the 2D mode and the integration is performed through a surface integral, representing the net electric flux flowing through the metal film of cross-section area $A_m$.

The plot in Fig. S4(b) shows the dispersion characteristics of the 2D CHPW proposed in this work. Obviously, the propagation loss is optimized when the flux is minimized in the metal layer, which corresponds to $t_{\varepsilon4} = 185\mathrm{nm}$. In this case, the normalized flux (0.003 V·m/W) can be rendered significantly smaller compared to $t_{\varepsilon4} = 170\mathrm{nm}$ (1.03 V·m/W). Such a sizable reduction in turn leads to improvement in the propagation loss, clearly demonstrating the effect of flux engineering on the propagation loss by diminishing the amount of electric field contained within the metal.

It should be noteworthy to point out that unlike the 1D counterpart, for 2D CHPW the regime of the field cancellation (zero-crossing) is curved within the metal layer as shown in Fig. S4(a). Consequently, the position of the zero-crossing is unable to predict the optimized waveguide geometry of 2D CHPW. Instead, the flux can indicate the balance of the positive and negative fields coexisting in the metal film, and hence the resulting strength of field that leads to the propagation loss. Since $E_z$ is the dominant field component for 2D CHPW, the optimization of the flux in $E_z$ thereby can lead to a minimum point in propagation loss for the long-range plasmonic mode using arbitrary material platform. Clearly, the electric flux in the metal offers a robust design metric that dictates the propagation loss of CHPW in 2D configuration, which in turn facilitates the design of our proposed plasmonic waveguide presented in this work.

**Modal Confinement**

While the sources of propagation loss have been determined, the effective mode area needs to be calculated in order to quantify the trade-off between propagation distance and mode confinement (34). However, conventional mode area calculation methods, such as area bounded by closed $1/e$ field magnitude contour relative to global field maximum, are inconsistent for most plasmonic modes as they do not follow Lorentzian field distributions found in typical TIR waveguides. As shown in Fig. S2, the plasmonic TM modes of the CHPW display extremely intense mode localization and sharp discontinuous sub-wavelength features across the layer interfaces. In order to quantify local field confinement and its role in light-matter interaction processes such as the Purcell effect, a phenomenological rather than a statistical measure is more suitable (35). The effective mode area A of the CHPW is therefore defined as ratio of the total mode energy density per unit length and peak energy density:

$$A = \frac{1}{\max\{W(\mathbf{r})\}} \int_{A_\infty} W(\mathbf{r}) \mathrm{d}A$$

$$W(\mathbf{r}) = \frac{1}{2} \mathrm{Re}\left\{\frac{\mathrm{d}[\omega\varepsilon(\mathbf{r})]}{\mathrm{d}\omega}\right\} |\mathbf{E}(\mathbf{r})|^2 + \frac{1}{2}\mu_0|\mathbf{H}(\mathbf{r})|^2 |$$

(S3)

where $W(\mathbf{r})$ is the mode energy density.

In Fig. S5(b-c), the total propagation loss per unit length and modal confinement are plotted against $\varepsilon_4$ thickness. It is observed that the mode area increases rapidly as thickness is decreased due to field leaking into the air cladding while it remains relatively constant at larger thicknesses as most of the field is localized in the low-index layer. At the optimized $\varepsilon_4$ thickness of 185nm, the propagation loss is 0.02 dB/μm with an effective mode area of $0.002\mu m^2$. In Table S1, the cross-section dimensions, propagation losses and mode areas for various short-range and long-range waveguides are compared against the CHPW. It is shown that relative to other guided plasmonic modes, the CHPW can achieve mode confinements similar to those found in plasmonic slot waveguides without sacrificing propagation length. The table also places dielectric TIR silicon nanowires in perspective to show that while extremely long-range propagation can be achieved due to their inherent lossless nature, TIR modes are comparatively loosely confined with mode areas much larger than SPP-based waveguides, thereby increasing the required light-matter interaction length significantly. Therefore, a solution that can integrate both conventional TIR waveguides for long-range propagation and low-insertion loss local CHPW devices is ideal.

**Coupling Between CHPW and Silicon Nanowires**

Integration of the CHPW into conventional silicon photonics allows designing of very compact light-matter devices using the former and chip-wide long-range propagation using the latter, but requires an efficient in-and-out coupling mechanism between the two platforms. The energy coupling between two dissimilar waveguiding systems is determined by momentum mismatch and spatial mode overlap. In the CHPW, the LR supermode is confined strongly near the metal interfaces, with evanescent tails extending into substrate and cladding, while silicon nanowires support TE and TM TIR modes confined strongly in the high index core. In Fig. S6, the fundamental modes of a 220nm-thick rectangular SOI nanowire are placed in perspective with the CHPW symmetric supermode. The dispersion plots and spatial mode overlap are also plotted as function of silicon nanowire width in order to find the optimized coupling conditions.

The silicon nanowire supports guided TE and TM modes above cut-off widths of 280nm and 320nm respectively. $TE_0$ is confined predominantly in the core, while $TM_0$ extends significantly to the oxide substrate and cladding due to reduced thickness and index contrast in the vertical direction. Moreover, since the core index is identical to the Si substrate beneath the buried oxide (BOX), these are not truly guided but rather quasi-leaky modes with leakage losses that decrease exponentially with BOX thickness. For a 2μm-thick BOX, substrate leakage of the $TE_0$ mode is negligible, but for $TM_0$ it is in the order of $10^1$ dB/mm for a width of 350nm and improves to $10^{-3}$ dB/mm at 800 nm (37). The vertical asymmetry between the oxide and air cladding also introduces mode hybridization of $TM_0$ and $TE_1$ due to structural birefringence (38). At widths 620-680nm, polarization of the eigenmodes cannot be distinguished and introduces undesired mode conversion.

For $TE_0$, direction of the momentum vector is fully mismatched as LR supermode is predominantly TM and consequently, mode overlap is zero. For $TM_0$, the phase is matched for a 1μm-wide SOI nanowire while spatial mode overlap peaks at 77% for a width of 720nm. These point to a relatively efficient end-fire coupling scheme in dimensions above the hybridization region. However, width of the SOI nanowire should be kept below 1μm to maintain single-mode operation in the TM polarization. In Fig. S7, the end-fire coupling efficiency into LR supermode and power lost to other modes are plotted from results obtained using Eigenmode Expansion (EME) solver provided by Lumerical. Results suggest a peak coupling efficiency in and out of the CHPW above 71% for an SOI nanowire width range of 700-850nm, or 1.5 dB insertion loss over a wide wavelength range. This indicates that highly efficient coupling into CHPWs can be achieved on the same substrate and enables simultaneous utilization of the two types of waveguides under the same technology process. Power transfer at the coupling interface is non-resonant and instantaneous with the smallest device footprint possible without additional mode conversion structures such as tapers or stubs.

**Optical Characterization Setup and Loss Measurements**

Measurements of CHPW propagation losses and ring resonator spectra were performed using the optical characterization setup shown in Fig. S8. To verify the broadband performance, the samples were designed to be facet-coupled. However, due to low coupling efficiency as a result of spatial mismatch between input fiber mode and on-chip nanowires, the input CW laser is first amplified using an erbium-doped fiber amplifier (EDFA). For mode imaging and propagation loss measurement, samples of CHPW with various top-side high-index layer thicknesses were fabricated. CHPW of different lengths were integrated into silicon nanowires in order to extract their losses using the cut-back method.

To extract the loss per unit length of the CHPWs and coupling efficiency of the end-butt junctions, the cut-back method is applied by measuring the linear drop in power for waveguides of increasing length via Eq. S4.

$$P_{out} = [C^2 \times e^{-\alpha L}] \times P_{in}$$

$$\ln\left(\frac{P_{out}}{P_{in}}\right) = -\alpha L + \ln C^2 \qquad \text{(S4)}$$

where $C$ is the junction coupling efficiency between silicon nanowires and CHPWs, $\alpha$ is the attenuation factor and $L$ is the CHPW section length.

The fabricated samples are characterized through standard end-fire coupling setup, where input light at $\lambda = 1550$ nm is coupled into the sample via a 60x objective lens and output light is collected with a 20x lens. In Fig. S9, the output mode profiles of samples with $t_{t4} = 190$nm and varying CHPW lengths are displayed under identical camera sensitivity and contrast settings. The output intensity decreases as the length of the CHPW increases, further confirming that the plasmonic waveguides and end-butt couplers are functional.

The power measurements at $\lambda = 1550$ nm and linear curve fits on CHPW samples with different $\alpha$-Si layer thicknesses ($t_{t4}$) are plotted in Fig. S10. The measured waveguide loss, coupling efficiency, and R-square of the linear fit are also shown on the graphs. The quality of the CHPW sets can be inferred from $R^2$ in curve fits, for which a value that is close to 1 suggests that the insertion losses increase linearly as function of $L$. This indicates that dominant propagation losses indeed come from absorption $\alpha$.

The loss trends observed as function of top-layer thickness corresponds and experimentally confirms the field-matching analysis performed in previous sections. The samples with $t_{t4} = 190$nm showed an average minimal propagation loss of 0.03 dB/µm, which is close to the value obtained from simulations. As the thickness is varied away from the optimal point, the losses can increase up to 0.07 dB/µm on the thinner side and 0.09 dB/µm when thicker. Deviations from simulated values can be attributed to fabrication such as roughness, as well as imprecise SEM imaging. In addition to propagation losses, the junction coupling efficiencies between silicon nanowires and CHPW were also extracted. As it is impossible to decouple input and output junction coupling in this scheme, they are assumed to be identical based on FDTD simulations. The experimental results indicate coupling efficiencies of 56-71%.

**CHPW Ring Resonators**

Ring resonators are narrow-band optical devices with the ability to increase the effective interaction length within a compact footprint. Common applications include optical filters (39), wavelength demultiplexing (40), dispersion compensators (41), modulators (42), laser generation (43), and optical signal processing (44). However, construction plasmonic waveguide-based ring resonators is challenging as a result of high intrinsic absorption losses, leading to low quality factors. As shown previously, the CHPW platform offers a significantly improved trade-off between loss and confinement, allowing extremely high light-matter interaction within a small mode volume with long effective propagation length. In this section, we evaluate CHPW ring resonators using semi-analytical models and verify the performance experimentally.

To model the directional coupling section, we use a generic model based on CMT and transfer matrix technique (TMT) for curved rectangular waveguides (45). Under this model, the coupling between two bent waveguides is described as the sum of coupling sections with infinitesimal length. Equation S5 models the power coupling ratio $\kappa$ between straight and bent waveguides, satisfying the relation $\kappa^2 + t^2 = 1$, where $t$ is the through transmission ratio.

$$\kappa = \sin\left[ R \int_{-\pi/2}^{\pi/2} K_\parallel(\theta)\cos^2\theta\, \mathrm{d}\theta \right]$$
$$K_\parallel(\theta) = \frac{2\gamma_{1x}^2 \gamma_{2x}^2}{\beta k_0^2 (n_1^2 - n_2^2)(2 + \gamma_{2x}a)} e^{-\gamma_{2x}(d + 2R\sin[2](\theta/2))}$$

$$(S5)$$

Where $K_\parallel(\theta)$ represents the coupling coefficient for each infinitesimal section with angle $\theta$ between the bent and straight waveguides. $R$ is the radius of curvature of the bent waveguide, $a$ is the core width, and $d$ is distance between the two waveguides at their minimum separation. $n_1$ and $n_2$ are the refractive indices of the core and cladding respectively, $\beta$ is the propagation constant, $k_0$ is the free space wavenumber, and $\gamma_{1x}$ and $\gamma_{2x}$ satisfy the relations $\gamma_{1x}^2 = k_0^2 n_1^2 - \beta^2$ and $\gamma_{2x}^2 = k_0^2 n_1^2 - k_0^2 n_2^2 - \gamma_{1x}^2$. To approximate the core index, $n_1$ is taken as the effective index of the CHPW with infinite width. In Fig. S11(a), the coupling ratios for bent waveguides are shown for radii of curvature from 1.0μm to 3.0μm.

To achieve critical coupling and maximize extinction ratio, the coupled power should be equal to the loss per round-trip inside the ring cavity. The optimal loss per round-trip for the CHPW was determined to be 0.8 dB for ring radius between 2.0 and 2.5μm, corresponding to a minimum gap separation of 285nm for critical coupling. The transmission spectra for all-pass CHPW ring resonators of 2.48μm ring radius were extracted using Lumerical FDTD and shown in Fig. S11(b). The simulation results indicate critical coupling at 285nm as well, with 30 dB extinction ratio and full-width half maximum (FWHM) of 2nm, corresponding to a theoretical quality factor of 775.

In Fig. S12, SEM of an all-pass CHPW ring resonator with 2.5μm is shown. Incoming signals are carried in from a silicon nanowire into a 10μm-long straight CHPW section coupled to the ring cavity, and carried out back into an output silicon nanowire. The transmission spectra for minimum gap distances of 270nm, 285nm and 300nm are compared. The maximum extinction ratio observed was 29 dB when the ring is critically coupled. However, the critical coupling condition is observed at 270nm instead of 285nm calculated in the semi-analytical models. This can be attributed to unaccounted propagation losses as a result of fabrication, such as sidewall roughness, layer interface roughness and absorption from the deposited amorphous silicon layer. The highest quality factor measured was 320 for ring resonators with 4.8nm FWHM. From Fig. S12(d), the CHPW ring resonators display an average 48pm/°C wavelength redshift as devices are heated from 20°C to 50°C.

The results obtained in CHPW ring resonators are favorable for design of whispering-gallery mode lasing applications on this platform (30). The optimized absorption loss contributes to relatively high Q-factors while maintaining tight mode confinement. Given the calculated 0.002μm$^2$ cross-sectional mode area, a 2.5μm-radius ring resonator only has an effective mode volume ($V_{eff}$) of 0.032μm$^3$. At a lasing wavelength band centered at 1550nm, our experimental device can achieve a normalized Purcell factor ($Q/V_{eff}$) of $1.5 \times 10^4$. In future work, we attempt to further enhance the Purcell factor via submicron ring radius to achieve even smaller mode volumes while at the same time curtailing bend losses as a result of sharper bends.

**References and Notes**

1. H. Raether, *Surface plasmons on smooth and rough surfaces and on gratings* (Springer-Verlag Berlin Heidelberg, ed. 1, 1988).

2. D. K. Gramotnev, S. I. Bozhevolnyi, Plasmonics beyond the diffraction limit. *Nature Photonics* **4**, 83-91 (2010).

3. S. Enoch, N. Bonod. *Plasmonics: From basics to advanced topics* (Springer-Verlag Berlin Heidelberg, ed. 1, 2012).

4. M. Kauranen, A. V. Zayats, Nonlinear plasmonics. *Nature Photonics* **6**, 737-748 (2010).

5. M. Stockman, Nanoplasmonics: past, present, and glimpse into future. *Opt. Express* **19**, 22029-22106 (2011).

6. S. A. Maier, *Plasmonics: Fundamentals and applications* (Springer-Verlag US, ed. 1, 2007).

7. W. L. Barnes, A. Dereux, T. W. Ebbesen, Surface plasmon subwavelength optics. *Nature* **424,** 824-830 (2003).

8. P. Berini, Figures of merit for surface plasmon waveguides. *Optics Express* **14**, 13030-13042 (2006).

9. R. Zia, J. A. Schuller, A. Chandran, M. L. Brongersma, Plasmonics: the next chip-scale technology. *Materials Today* **9**, 20-27 (2006).

10. M. Fleischmann, P. J. Hendra, A. J. McQuillan, Raman spectra of pyridine adsorbed at a silver electrode. *Chemical Physics Letters* **26**, 163-166 (1974).

11. S. I. Bozhevolnyi, V. S. Volkov, E. Devaux, J. Y. Laluet, T.W. Ebbesen, Channel plasmon subwavelength waveguide components including interferometers and ring resonators. *Nature* **440**, 508-511 (2006).

12. T. Wu, Y. Liu, Z. Yu, Y. Peng, C. Shu, C., H. Ye, The sensing characteristics of plasmonic waveguide with a ring resonator. *Optics Express* **22**, 7669-7677 (2006).

13. G. Li, C. Sterke, S. Palomba, Figure of merit for kerr nonlinear plasmonic waveguides. *Laser & Photonics Reviews* **10**, 639-646 (2016).

14. J. B. Khurgin, G. Sun, Practicality of compensating the loss in the plasmonic waveguides using semiconductor gain medium. *Appl. Phys. Lett.* **100**, 011105 (2012).

15. P. Berini, Long-range surface plasmon polaritons. *Advances in Optics and Photonics* **1**, 484-588 (2009).

16. M. Z. Alam, J. Meier, J. S. Aitchison, M. Mojahedi, Super mode propagation in low index medium. *CLEO/QELS*, page JThD112 (2007).

17. R. F. Oulton, V. J. Sorger, D. A. Genov, D. F. P. Pile, X. Zhang, A hybrid plasmonic waveguide for subwavelength confinement and long-range propagation. *Nature Photonics* **2**, 496 (2008).

18. W. Ma, A. S. Helmy, Asymmetric long-range hybrid-plasmonic modes in asymmetric nanometer-scale structures. *Journal of the Optical Society of America B* **31**, 1723-1729 (2014).

19. C. Lin, H. Wong, B. Lau, M. Swillam, A. S. Helmy, Efficient broadband energy transfer via momentum matching at hybrid junctions of guided-waves. *Applied Physics Letters* **101**, 123115 (2012).

20. S. Zhu, G. Lo, D. Kwong, Submicron-radius plasmonic racetrack resonators in metal-dielectric-Si hybrid plasmonic waveguides. *IEEE Photonics Technology Letters* **26**, 833-836 (2014).

21. S. Zhu, G. Lo, D. Kwong, Performance of ultracompact copper-capped silicon hybrid plasmonic waveguide-ring resonators at telecom wavelengths. *Optics Express* **20**, 15232-15246 (2012).

22. T. Holmgaard, Z. Chen, S. Bozhevolnyi, L. Markey, A. Dereux, A. Krasavin, A. Zayats, Wavelength selection by dielectric-loaded plasmonic components. *Applied Physics Letters* **94**, 051111 (2009).

23. A. Biberman, M. J. Shaw, E. Timurdogan, J. B. Wright, M. R. Watts, Ultralow-loss silicon ring resonators. *Optics Letters* **37**, 4236-4238 (2012).

24. C. Lin, A. S. Helmy, Dynamically reconfigurable nanoscale modulators utilizing coupled hybrid plasmonics. *Scientific Reports* **5**, 12313 (2015).

25. Y. Su, C. Lin, P. Chang, A. S. Helmy, Highly sensitive wavelength-scale amorphous hybrid plasmonic detectors. *Optica* **4**, 1259-1262 (2017).

26. A. V., Krasavin, S. Randhawa, J. S. Bouillard, J. Renger, R. Quidant, A. V. Zayats, Optically-programmable nonlinear photonic component for dielectric-loaded plasmonic circuitry. *Optics Letters* **19**, 25222-25229 (2011).

27. S. Randhawa, S. Lacheze, J. Renger, A. Bouhelier, R. de Lamaestre, A. Dereux, R. Quidant, Performance of electro-optical plasmonic ring resonators at telecom wavelengths. *Optics Express* **20**, 2354-2362 (2012).

28. M. Mesch, B. Metzger, M. Hentschel, H. Giessen, Nonlinear plasmonic sensing. *Nano Letters* **16**, 3155-3159 (2016).

29. R. F. Oulton, V. J. Sorger, T. Zentgraf, R. M. Ma, C. Gladden, L. Dai, G. Bartal, X. Zhang, Plasmon lasers at deep subwavelength scale. *Nature* **461**, 629-632 (2009).

30. C. Xiang, C. K. Chan, J. Wang, Proposal and numerical study of ultra-compact active hybrid plasmonic resonator for sub-wavelength lasing applications. *Scientific Reports* **4**, 3720 (2014).

31. T. Hümmer, F. J. Garcia-Vidal, L. Martin-Moreno, D. Zueco, Weak and strong coupling regimes in plasmonic-GED. *Physical Review B* **87**, 115419 (2013).

32. J. A. Dionne, L. A. Sweatlock, H. A. Atwater, A. Polman, Plasmon slot waveguides: Towards chip-scale propagation with subwavelength-scale localization. *Physical Review B* **73**, 035407 (2006).

33. J. J. Burke, G. I. Stegeman, T. Tamir, Surface-polariton-like waves guided by thin, lossy metal films. *Physical Review B*, **33**, 5186-5201 (1986).

34. R. Buckley, P. Berini, Figures of merit for 2d surface plasmon waveguides and application to metal stripes. *Optics Express* **15**, 12174-12182 (2007).

35. R. F. Oulton, G. Bartal, D. F. P. Pile, X. Zhang, Confinement and propagation characteristics of subwavelength plasmonic modes. *New Journal of Physics* **10**, 105018 (2008).

36. Z. Sheng, C. Qiu, H. Li, L. Li, A. Pang, A. Wu, X. Wang, S. Zou, F. Gan, Low loss silicon nanowire waveguide fabricated with 0.13μm CMOS technology. *Optical Society of America*, page ATh1B.2 (2012).

37. P. Dumon, Ultra-compact integrated optical filters in silicon-on-insulator by means of wafer-scale technology (2007). http://lib.ugent.be/catalog/rug01:001197413

38. D. Dai, J. E. Bowers, Novel concept for ultracompact polarization splitter-rotator based on silicon nanowires. *Optics Express* **19**, 10940-10949 (2011).

39. D. G. Rabus, *Integrated ring resonators: The compendium* (Springer-Verlag Berlin Heidelberg, ed. 1, 2007).

40. M. S. Rasras, K. Y. Tu, D. M. Gill, Y. K. Chen, A. E. White, S. S. Patel, A. Pomerene, D. Carothers, J. Beattie, M. Beals, J. Michel, L. C. Kimerling, Demonstration of a tunable microwave-photonic notch filter using low-loss silicon ring resonators. *Journal of Lightwave Technology* **27**, 2105-2110 (2009).

41. C. K. Madsen, J. A. Walker, J. E. Ford, K. W. Goossen, T. N. Nielsen, G. Lenz, A tunable dispersion compensating mems all-pass filter. *IEEE Photonics Technology Letters* **12**, 651-653 (2000).

42. H. W. Chen, A. W. Fang, J. Bovington, J. D. Peters, J. E. Bowers, Hybrid silicon tunable filter based on a Mach-Zehnder interferometer and ring resonator. *2009 International Topical Meeting on Microwave Photonics*, pages 1-4 (2009).

43. D. J. Moss, R. Morandotti, A. L. Gaeta, M. Lipson, New CMOS-compatible platforms based on silicon nitride and hydex for nonlinear optics. *Nature Photonics* **7**, 597-607 (2013).

44. A. Fushimi, T. Tanabe, All-optical logic gate operating with single wavelength. *Optics Express* **22**, 4466-4479 (2014).

45. C. S. Ma, X. Yan, Y. Z. Xu, Z. L. Qin, X. Y. Wang, Characteristic analysis of bending coupling between two optical waveguides. *Optical and Quantum Electronics* **37**, 1055-1067 (2005).

## Acknowledgments

**Funding:** This project was funded by the Natural Sciences and Engineering Research Council (NSERC) of Canada.

**Author contributions:** All authors contributed equally to this project.

**Competing interests:** The authors declare that there are no competing interests.

## Figures and Tables

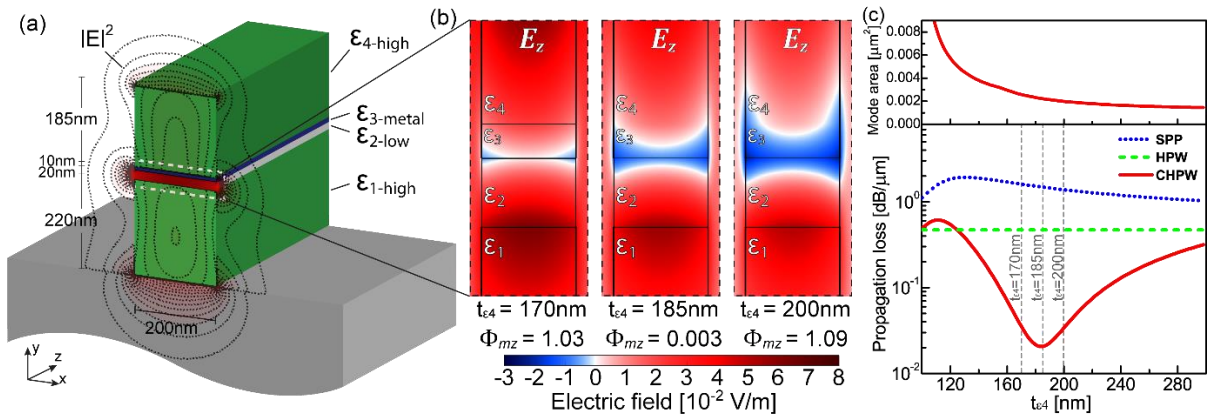

**Fig. 1. Schematic and modal properties of the CHPW. (A)** A schematic of the composite hybrid plasmonic waveguide core cross-section discussed in this work. This CHPW is a 4-layer stack that consists of 220nm bottom high-index Si ($\varepsilon_1$), 20 low-index SiO$_2$ spacer ($\varepsilon_2$), 10nm Al ($\varepsilon_3$) metal layer, and 185nm top high-index α-Si ($\varepsilon_4$). The long-range mode in this structure is primarily confined within the low-index layer $\varepsilon_2$, as shown in the overlaid modal E-field intensity profile. **(B)** The cross-section area around the $\varepsilon_3$ metal layer is expanded to plot the longitudinal $E_z$ field profile. By varying the $\varepsilon_4$ thickness, the longitudinal electric flux $\Phi_{mz}$ (in V·m/W units) can be minimized by engineering the $E_z$ within the metal. **(C)** The calculated modal area and propagation loss of the long-range mode as function of the layer $\varepsilon_4$ thickness. The thickness for optimal propagation loss ($t_{\varepsilon4}$ = 185nm) corresponds to minimum flux within the metal layer. Using this method, the propagation losses of modes in CHPWs can be drastically reduced for a wide range of structures, without any restrictions on symmetry, while maintaining an extremely localized effective mode area of 0.002μm$^2$.

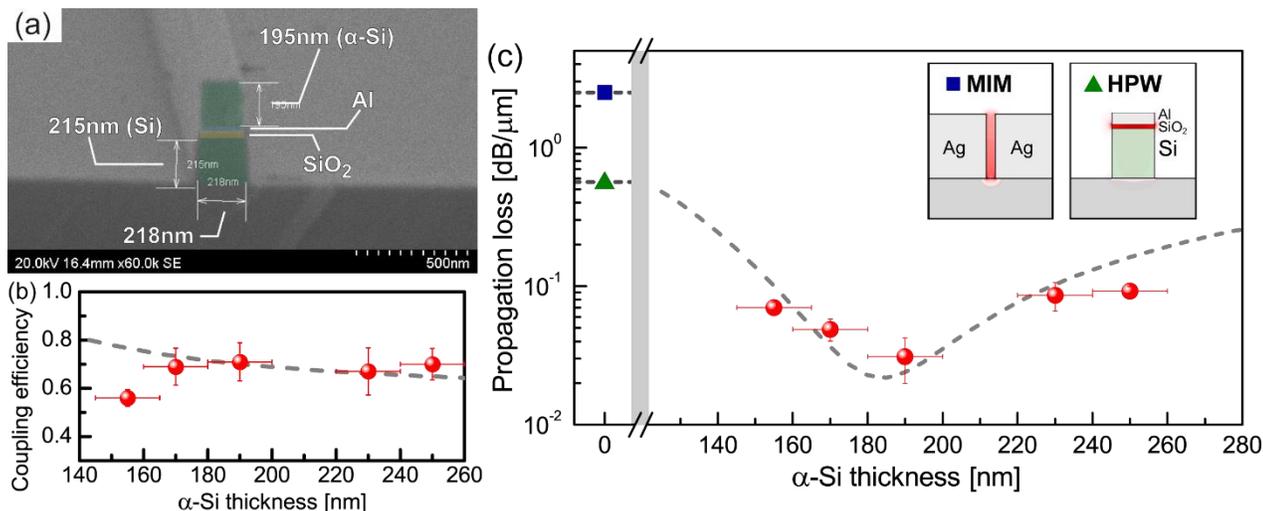

**Fig. 2. Experimental loss measurements. (A)** Scanning-electron micrograph of a cleaved CHPW facet cross-section showing small fabrication variations in geometry compared to the simulated structure. In test devices, CHPW samples with different α-Si thicknesses are excited via end-butt coupling in and out of dielectric silicon nanowires. **(B)** The plot shows the measured coupling efficiency between a junction formed by an 800nm-wide silicon nanowire and a 200nm-wide CHPW, compared to theoretical values obtained via FDTD modeling (gray dashed line). **(C)** The plot shows propagation loss measured using cut-back method for samples with different α-Si thickness, compared to theoretical values obtained from FDTD modeling (gray dashed line). For reference, the measured propagation losses of a silver-air-silver plasmonic slot waveguide and single-sided HPW that does not have a top α-Si layer are also compared.

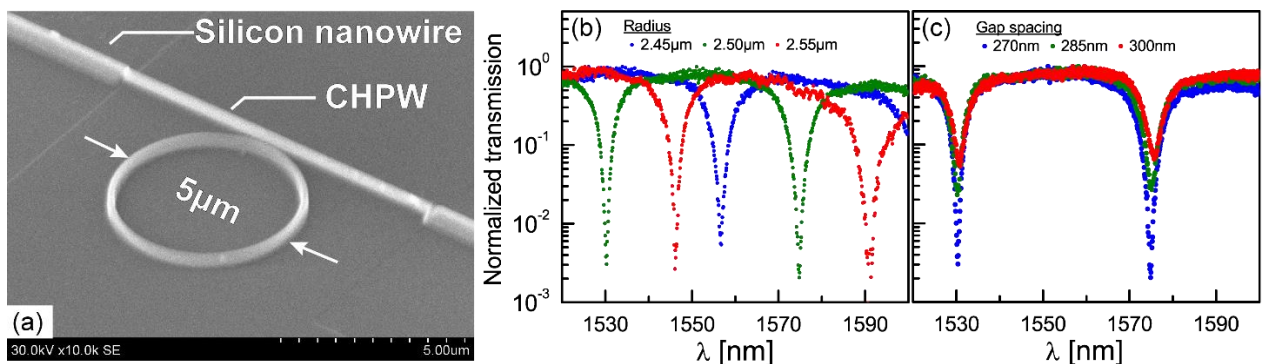

**Fig. 3. CHPW microring resonators. (A)** Bird's eye view in scanning-electron micrograph of a 2.5μm-radius CHPW microring resonator in an all-pass configuration. The input and output light is interfaced with the CHPW section utilizing 800nm dielectric silicon nanowires. **(B)** Measured transmission spectra of CHPW ring resonators as ring radius is varied by 50nm. The critically-coupled operating point corresponded to a gap spacing of 270nm. **(C)** The extinction ratio decreases due to under-coupling as the gap spacing increases for CHPW resonators with 2.5μm radius.

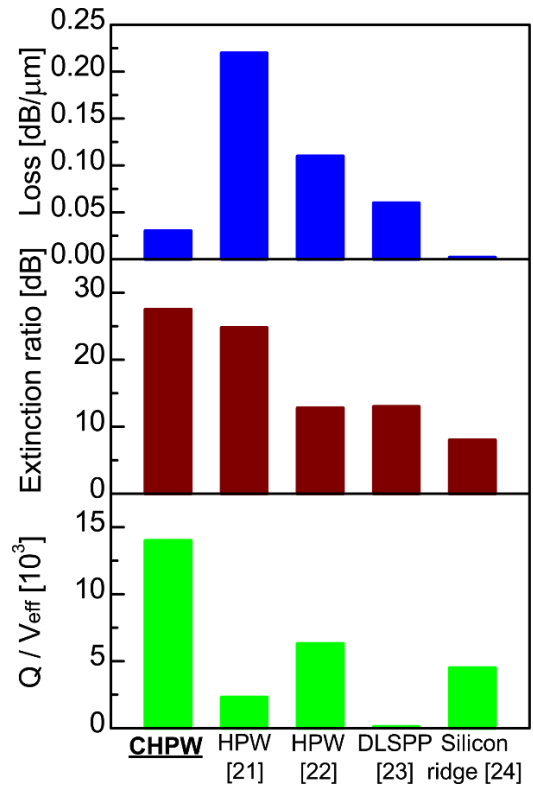

**Fig. 4. Purcell factor comparison of CHPW ring resonators against literature.** The measured propagation loss, ring resonator extinction ratio and normalized Purcell factor of the CHPW ring resonators are compared against representative, previously reported plasmonic and silicon ring resonators. In typical plasmonic waveguides, highly localized plasmonic modes are afflicted by low quality factors as a result of losses, while subwavelength confinement is not attainable in diffraction-limited dielectric waveguides. In contrast, the CHPW can achieve record low losses of 0.03 dB/μm through flux engineering while concurrently asserting a nanoscale modal confinement of only 0.002μm$^2$, leading to an effective mode volume of 0.032μm$^3$ for 2.5μm ring radius. Through this combination, $1.5 \times 10^4$ normalized Purcell factor can be achieved as the loss-confinement trade-off inherent in plasmonics is drastically alleviated using our structures.

## Supplementary Materials

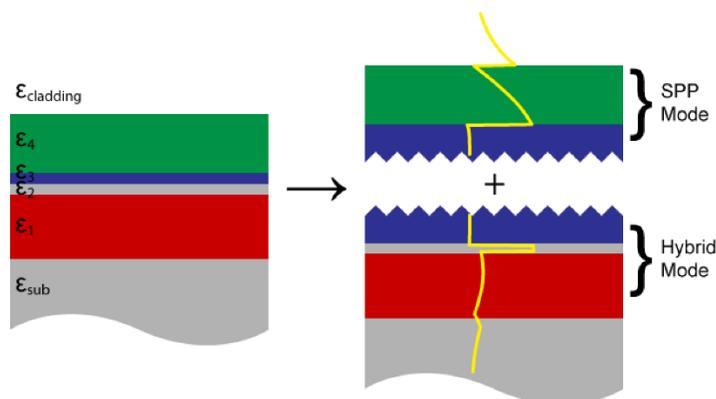

**Fig. S1. Coupling mechanism of the CHPW.** The CHPW can be treated as superposition of two individual decoupled HPW and SPP modes when the metal film is thicker than its skin depth.

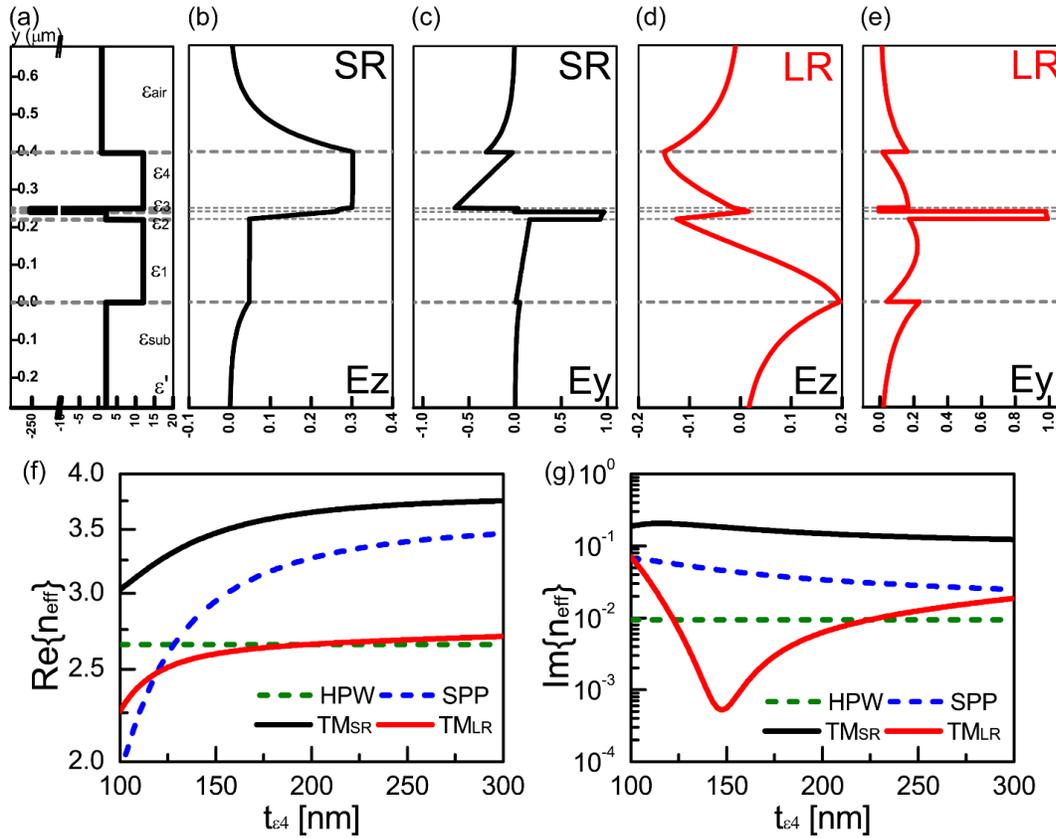

**Fig. S2. Plasmonic supermodes and mode properties of 1D CHPW.** (**A**) Material permittivity for each of the layers in the 1D CHPW multi-layer stack. (**B**) $E_z$ and (**C**) $E_y$ field components of the short-range plasmonic supermode. (**D**) $E_z$ and (**E**) $E_y$ field components of the long-range plasmonic supermode. (**F**) The propagation constant and (**G**) extinction coefficient of the coupled short-range and long-range TM supermodes, are compared against the originally decoupled HPW and SPP modes, as the thickness of the high-index $\varepsilon_4$ layer is varied. While the $E_y$ field components of either SR or LR modes show minimal field overlap with the $\varepsilon_3$ metal layer, the $E_z$ component of the SR mode strongly overlaps as evident in the three orders of magnitude difference in absorption loss.

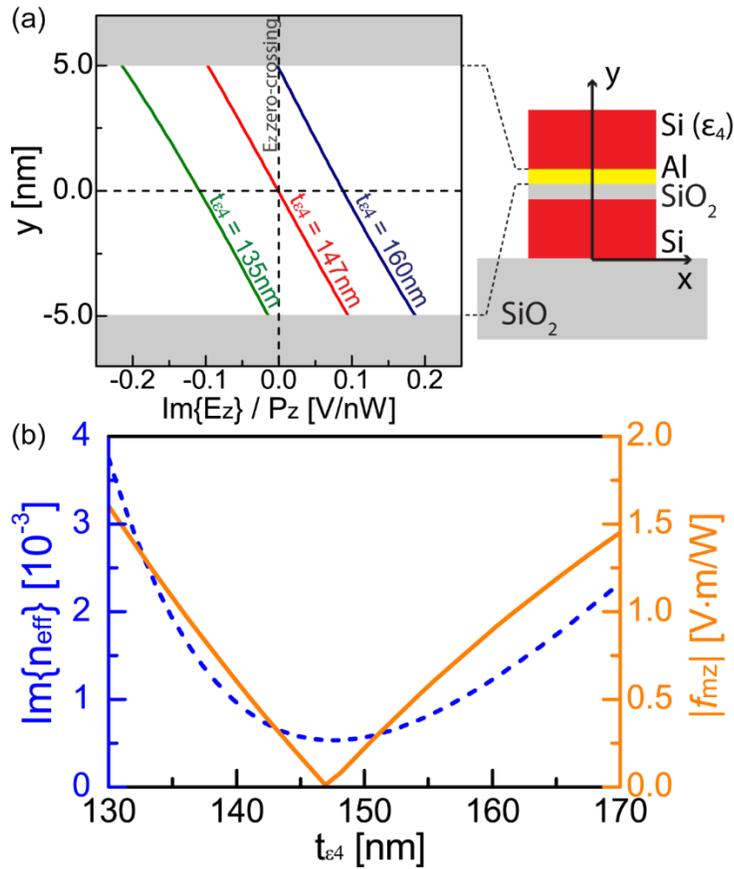

**Fig. S3. 1D Electric field distribution inside the metal layer.** (**A**) The normalized 1D field profile of CHPW within the Al metal layer of 1D CHPW stack is plotted for the optimal field cancellation at $t_{\varepsilon 4} = 147$nm and when it deviates at $t_{\varepsilon 4} = 135$ and 165nm. The $y$-axis is centered at the middle of the 10nm-thick metal layer while $x$-axis shows the anti-symmetric field profile centered at the zero-crossing. (**B**) The extinction coefficient (left) and normalized integration of $E_z$ profile (right) contained within the metal film of 1D CHPW as $t_{\varepsilon 4}$ varies. The minimum absorption corresponds to when $f_{mz}$ is closest to zero, indicating a balanced anti-symmetric distribution throughout the metal layer.

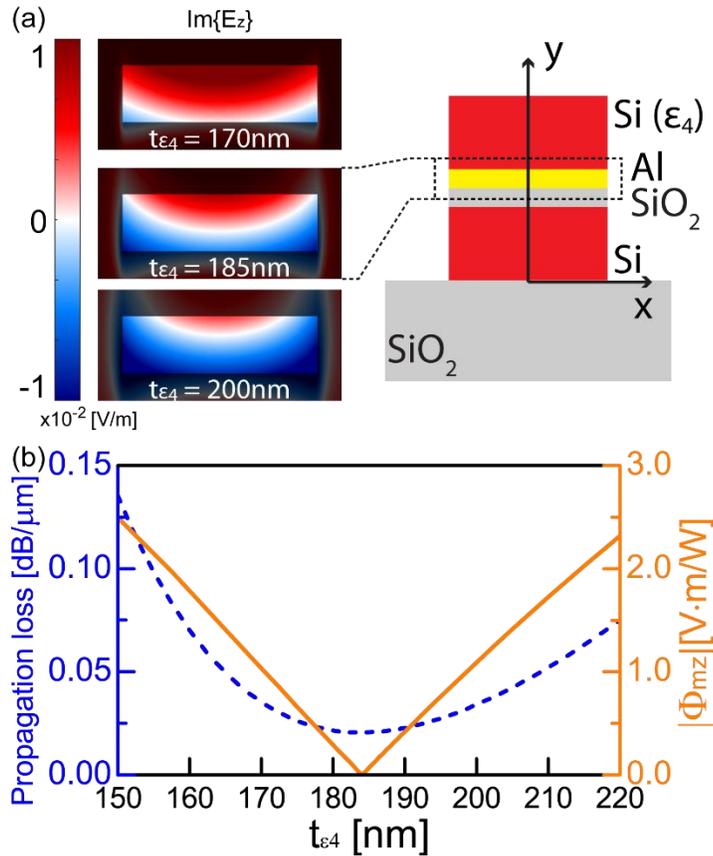

**Fig. S4. 2D Electric field distribution inside the metal layer. (A)** The cross-section area around the metal layer is expanded to plot the 2D $E_z$ field profiles at various $\varepsilon_4$ thicknesses, showing the anti-symmetric nature of the $E_z$ component. **(B)** The propagation loss (left) and total normalized electric flux $\Phi_{mz}$ contained within the metal film of CHPW as function of $\varepsilon_4$ thickness. To optimize the propagation losses, $t_{\varepsilon 4}$ is tuned until $\Phi_{mz}$ is closest to zero, indicating a uniformly balanced anti-symmetric field distribution throughout the metal layer as shown in (a).

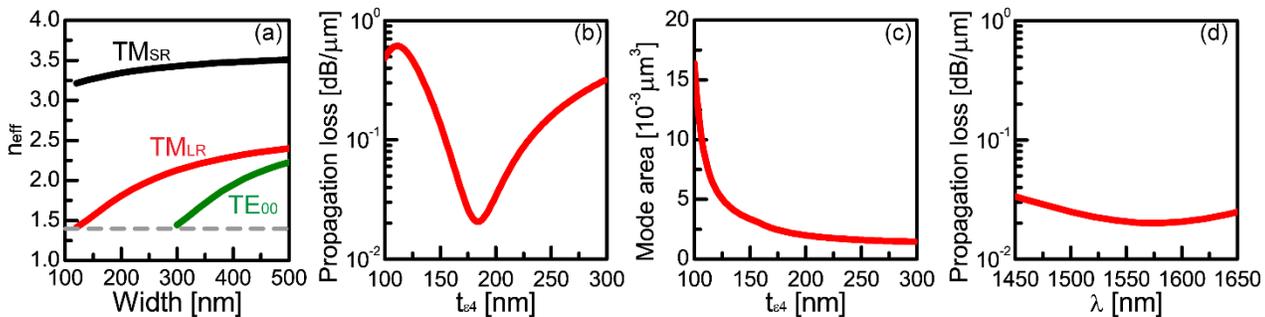

**Fig. S5. Mode properties of the 2D CHPW. (A)** Effective mode index of the CHPW plasmonic short-range and long-range supermodes, and the fundamental TE supermode as the width changes. The TE modes are completely cut-off at core widths below 300nm as the effective index drops below the substrate index. **(B)** The total propagation loss of the 2D 200nm-wide CHPW as $\varepsilon_4$ thickness varies and its **(C)** The effective mode area. At the optimized condition, the CHPW has a modal area of 0.002µm² and only varies by 5% if $t_{\varepsilon 4}$ deviates by 10%. The modal confinement worsens rapidly as $t_{\varepsilon 4}$ decreases due to field leaking into the air cladding. **(D)** The propagation loss does not vary significantly over a 200nm bandwidth centered at 1550nm, demonstrating that the low-loss condition does not strongly depend on wavelength.

| | Core cross-section [nm²] | $n_{eff}$ | Loss [dB/μm] | Mode area [μm²] |
|---|---|---|---|---|
| Si (TE) [10] | 500 × 220 | 2.41 | 0.24 × 10⁻³ | 0.13 |
| Si (TM) [10] | 500 × 220 | 1.60 | 0.06 × 10⁻³ | 0.17 |
| MIM [11] | 50 × 340 | 1.43 | 0.61 | 0.001 |
| HPW [12] | 200 × 252 | 2.28 | 0.22 | 0.02 |
| HPW [13] | 180 × 360 | 2.27 | 0.11 | 0.01 |
| **CHPW** | **200 × 435** | **1.81** | **0.02** | **0.002** |

**Table S1. Comparisons for various short-range and long-range waveguides against the CHPW.** The metrics compared are physical cross-section dimensions, propagation loss and mode area.

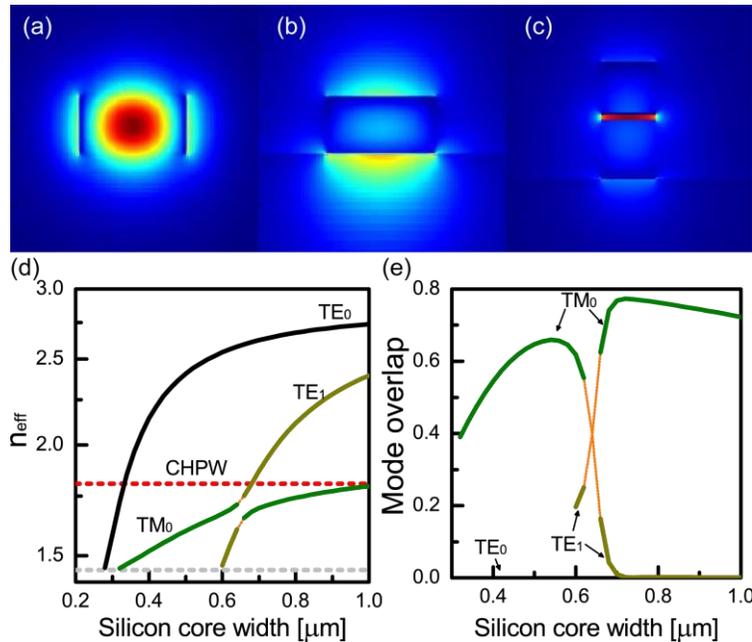

**Fig. S5. Mode profiles of silicon nanowires compared to CHPW.** The fundamental (**A**) TE and (**B**) TM modes supported by a 500nm-wide SOI nanowire and (**C**) LR supermode of the 200nm-wide CHPW. (**D**) Effective mode index of the SOI modes as width is varied from 200nm to 1μm, dashed gray line indicates the material index of SiO₂, for which modes with lower index are cut-off and leak into the substrate. (**E**) Spatial mode overlap between SOI modes to the LR supermode.

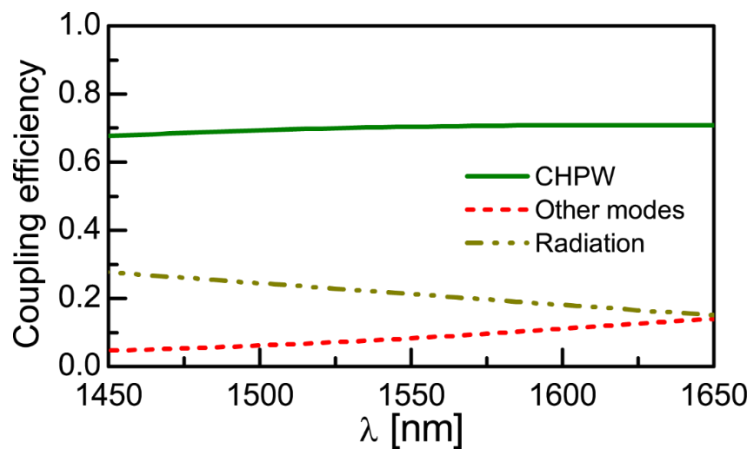

**Fig. S6. Coupling efficiency between CPHW and silicon nanowires.** Coupling efficiency into LR supermode and power lost to other modes over a 0.3µm wavelength bandwidth centered at 1550nm, for a TM-polarized 800nm-wide silicon nanowire mode.

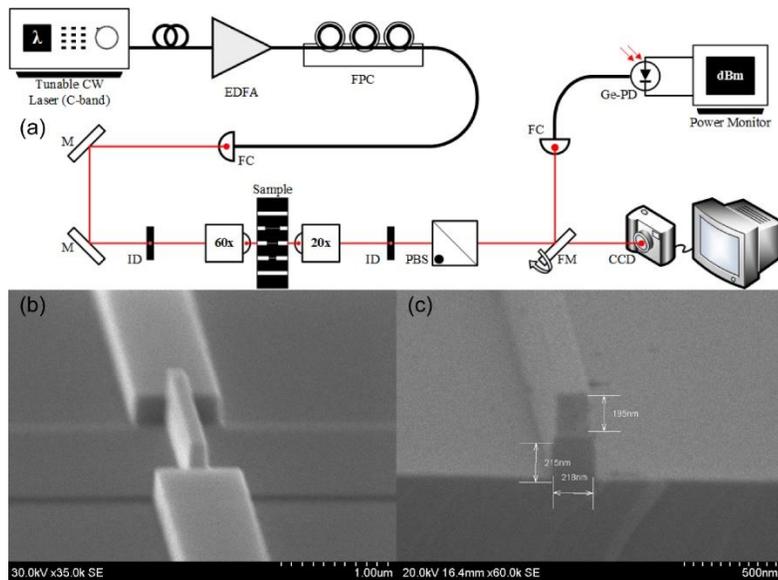

**Fig. S7. Characterization setup for experimental samples.** (**A**) Schematic of the optical characterization setup. CW: continuous wave; EDFA: erbium-doped fiber amplifier; FPC: fiber polarization controller; FC: fiber collimator; M: mirror; ID: iris diaphragm; PBS: polarization beam splitter; FM: flip-mount mirror; CCD: charged-coupled device camera; Ge-PD: germanium photodetector. (**B**) The sample consists of facet-coupled SOI nanowires with CHPW integrated via end-butt coupling. (**C**) A cleaved sample facet SEM of the CHPW cross-section of the 200nm-wide CHPW stack.

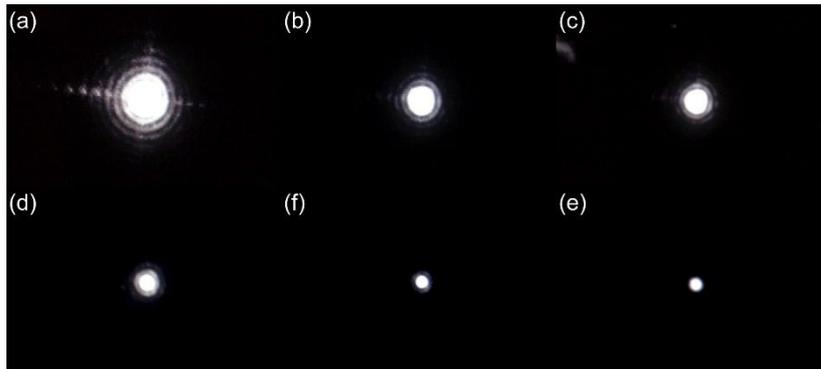

**Fig. S8. Observed optical mode profiles.** CCD camera captured mode profiles at λ = 1550nm of (**A**) 0.8µm-wide reference TM-mode SOI nanowire, and end-butt coupled CHPW with lengths of (**B**) 20µm, (**C**) 40µm, (**D**) 100µm, (**E**) 200µm and (**F**) 400µm.

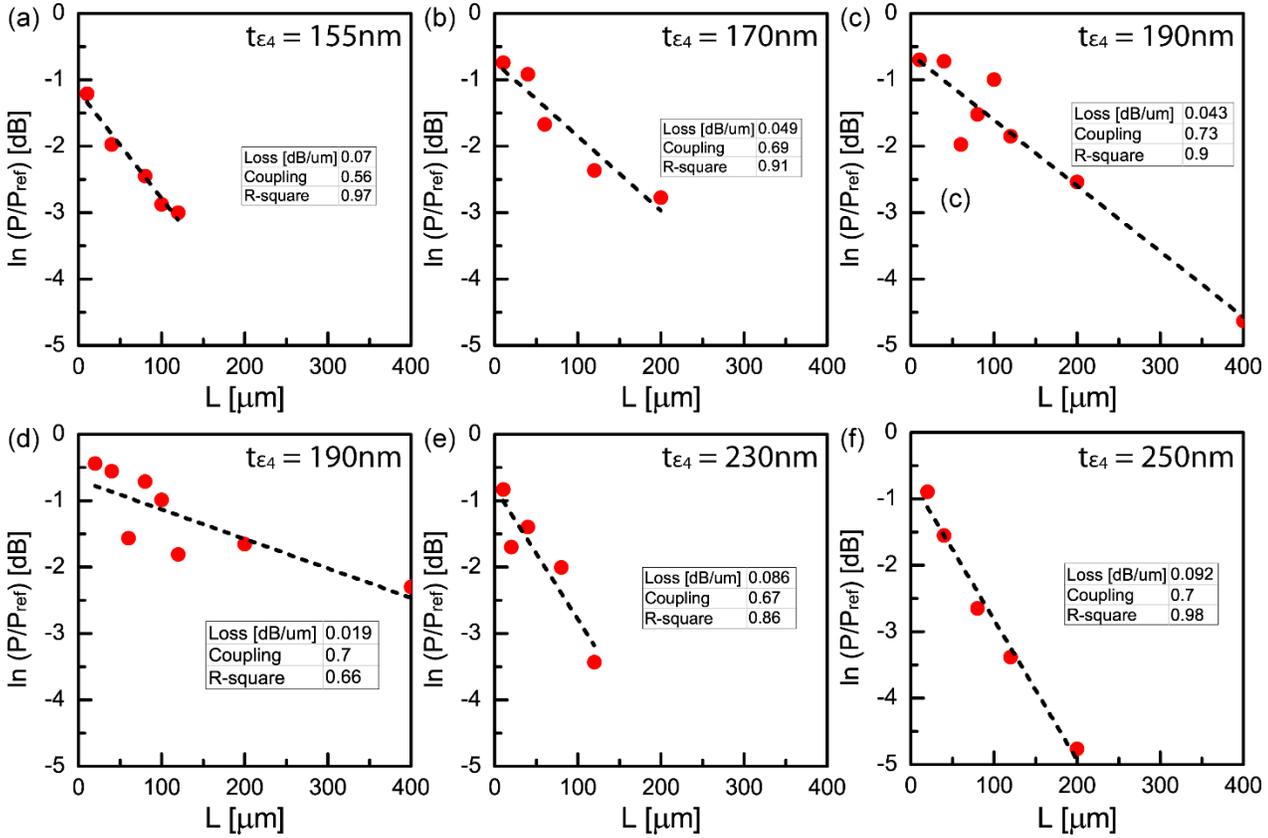

**Fig. S9. Propagation loss and coupling efficiency measurements.** Power transmission and linear curve fits for CHPWs of increasing length. The cut-back method is applied to extract loss and coupling efficiency for samples with varying $t_{\varepsilon4}$ thicknesses measured using SEM: (**A**) 155nm, (**B**) 170nm, (**C-D**) 190nm, (**E**) 230nm, (**F**) 250nm. The slope of each plot corresponds to the attenuation loss factor while the *y*-intercept indicates the coupling efficiency from a silicon nanowire, as described in Eq. S4.

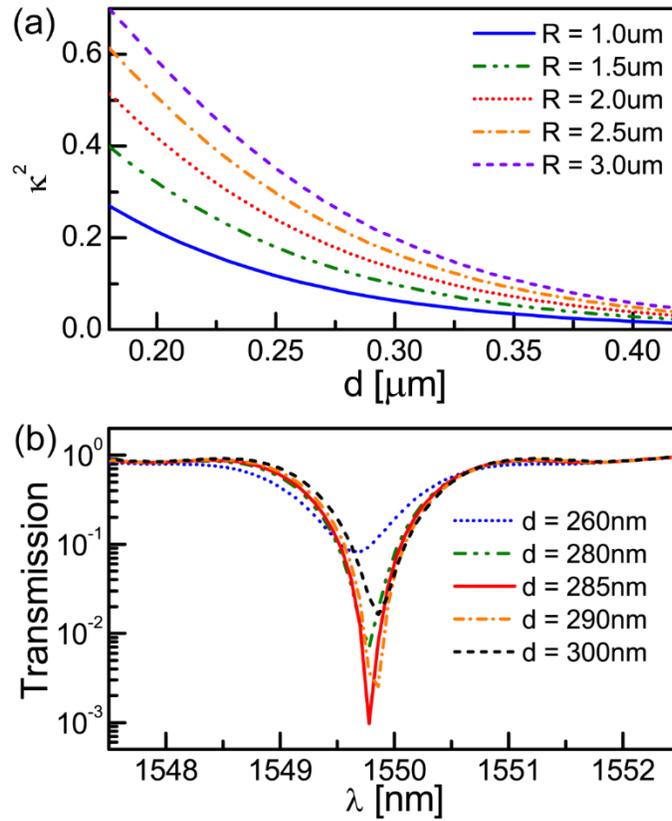

**Fig. S10. Semi-analytical modeling of CHPW ring resonators.** (**A**) Directional coupling ratio $\kappa^2$ between straight CHPW and bent section with various radii of curvature approximated using Eq. S5. (**B**) Theoretical transmission spectrum for an all-pass CHPW ring resonator with 2.48μm ring radius simulated using FDTD. The resonator is theoretically critically coupled at a gap separation of 285nm given an intrinsic 0.8 dB loss per round-trip.

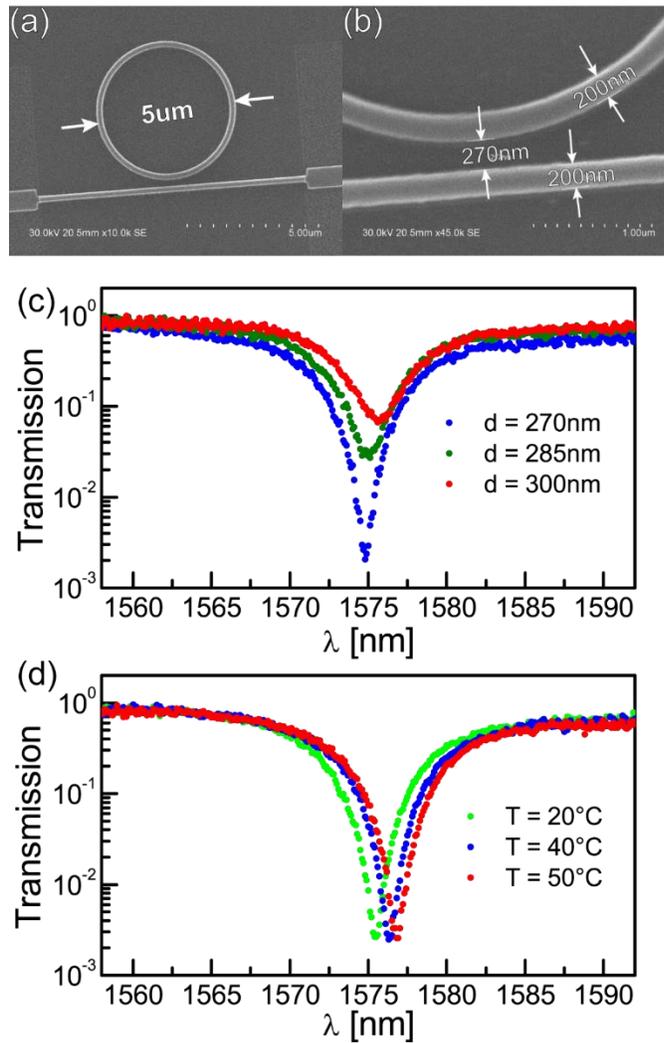

**Fig. S11. Experimental measurements of CHPW ring resonators. (A)** SEM image of an all-pass CHPW ring resonator with 2.5µm radius, integrated in and out of silicon nanowires, and **(B)** directional coupler section with minimum gap separation of approximately 270nm. **(C)** Transmission spectra experimentally measured for the all-pass CHPW ring resonators with ring radius of 2.5µm at various minimum gap distances. Compared to theoretical predictions, the critical coupling condition is observed at 270nm due to additional losses incurred from fabrication, while the FWHM is also wider leading to lower quality factors. **(D)** The ring resonators were subjected to temperature increase in a copper-plated heat stage and measured an average 48pm/°C wavelength redshift as temperature increases.